\documentclass[ preprint, %nolinenumbers,
 amsmath,amssymb,
 aps, physrev,
]{revtex4-2}
\usepackage{graphicx}% Include figure files
\usepackage{dcolumn}% Align table columns on decimal point
\usepackage{bm}% bold math
\begin{document}
%\preprint{APS/123-QED}
\title{\textbf{Capping layer dependent anti-correlation between magnetic damping and spin-orbital to charge conversion} 
}% 
\author{Antarjami Sahoo 1, Swayang Priya Mahanta 1, and Subhankar Bedanta* 1,2}
\affiliation{1 Laboratory for Nanomagnetism and Magnetic Materials (LNMM), School of Physical Sciences, National Institute of Science Education and Research (NISER), An OCC of Homi Bhabha National Institute (HBNI), Jatni, Odisha 752050, India}
\affiliation{2 Center for Interdisciplinary Sciences (CIS), National Institute of Science Education and Research (NISER), An OCC of Homi Bhabha National Institute (HBNI), Jatni, Odisha 752050, India}
\author{Subhankar Bedanta}%
 \email{sbedanta@niser.ac.in}
\date{\today}
\begin{abstract}
The magnetic Gilbert damping and spin-orbital to charge interconversion phenomenon play vital role in controlling the modern spintronics device performances. Though the ferromagnets (FMs) and heavy metals (HMs) are considered to be the key components of the future spin-orbit torque magnetic random access memory (SOT-MRAM) devices, recently the integration of lighter materials with low intrinsic spin-orbit coupling (SOC) in spintronics devices has proven to be noteworthy. Here we demonstrate the efficient control of magnetization dynamics of $\beta$-W/CoFeB bilayer when capped by low SOC organic and inorganic layers. The C$_{60}$ capping layer (CL) significantly enhances the magnetization relaxation process compared to the CuO$_x$ in $\beta$-W/CoFeB/CL heterostructures, while the static magnetic properties remain in-different irrespective of the nature of CL. Interestingly, the spin-orbital to charge conversion phenomenon is found to be enhanced for $\beta$-W/CoFeB/CuO$_x$ stacking compared to the $\beta$-W/CoFeB/C$_{60}$ heterostructure, signifying the anti-correlation between the magnetic damping and spin-orbital to charge conversion. The results are interpreted by the interfacial phenomena, like the orbital Rashba effect, two-magnon scattering, and interfacial spin memory loss. Our detailed experimental investigations shed light on the importance of low SOC materials in effectively tuning the magnetization dynamics for the development of future power efficient spintronics devices.  
\end{abstract}

\maketitle

%\tableofcontents

\section{\label{sec:level1}Introduction}
The usage of conventional complementary metal-oxide semiconductor (CMOS)-technology based microelectronics has shrunk in recent times due to the rise of big data-driven applications, like artificial intelligence (AI) and Internet of things (IOT). Subsequent development in the spintronics research domain has become pivotal at the dusk moment of Moore’s law. The spin transfer torque (STT) and spin-orbit torque (SOT) induced magnetization control forms the backbone of the spintronics technology which can overcome the shortcomings of the present microelectronics industry for advancement in the information technology sector \cite{han2022microelectronics, guo2021spintronics}. The STT-magnetic random access memory (STT-MRAM) has already been commercialized, whereas the capacity per chip has to be pushed into the GB range for the industrial application of SOT-MRAMs. The proto-type SOT-MRAMs have already shown higher endurance and faster writing speed as compared to the STT-MRAMs. The material parameters, like magnetic Gilbert damping, magnetic anisotropy, and saturation magnetization of ferromagnet (FM) become important for STT-based applications. Whereas, the spin current generation in the non-magnetic (NM) heavy metal (HM), efficient spin angular momentum transfer across the interface, and subsequent, sizable charge-to-spin interconversion in HM/FM/capping layer (CL) are the key requirements for the SOT based spintronics applications along with the other material parameters of ferromagnets. 
\par
Especially the light materials (i.e. materials with low spin-orbit coupling (SOC)) are usually employed as CLs. However, the materials with low intrinsic SOC, like Ti, Cr, Mn have recently garnered significant attention thanks to the discovery of orbital Hall effect (OHE) and orbital Rashba Edelstein effect (OREE) \cite{wang2023inverse, lyalin2023magneto, sala2023orbital, go2021orbital}. The Onsager reciprocal of these orbital angular momentum (OAM) mediated phenomena are known as inverse orbital Hall effect (IOHE) and inverse orbital Rashba Edelstien effect (IOREE) \cite{wang2023inverse, go2021orbital, sahoo2024efficient}. The OHE/ORE can lead to non-equilibrium accumulation of OAM and consequently, the orbital torque (OT) arising by virtue of OAM can alter the magnetization orientation of the adjacent FM layer \cite{sahoo2024efficient, santos2024exploring}. The generation of OT is not only limited to bulk materials and it can also arise at metal/oxide interface (like, Cu/Al$_2$O$_3$) and naturally oxidized Cu (CuO$_x$) surface \cite{kim2021nontrivial, sahoo2024efficient}. In particular, the hybridization between Cu $d$ states and O $p$ states can induce a chiral OAM texture in $k$-space, which consequently results in gigantic ORE \cite{go2021orbital}. On the other hand, the organic compounds also possess low SOC as they are composed of relatively lighter elements, like C, H, N etc. Sun et al., have shown a generation of inverse spin Hall effect (ISHE) in C$_{60}$ when they are stacked alongside of NiFe via pulsed ISHE technique \cite{sun2016inverse}. However, the experimental and theoretical research is limited when it comes to exhibition of SHE, OHE, or ORE in these types of compounds. 
\par
Moreover, CuO$_x$ has proven to be one of the powerful candidates when it comes to the generation of OT \cite{sahoo2024efficient, santos2024exploring}. At the same time, organic molecules can efficiently get hybridized with 3d metals (like Co, Fe) and 5d metals (like Pt, W) \cite{mallik2019enhanced, pandey2023perspective, sahoo2024molecular, alotibi2021enhanced}. Nevertheless, the effect of these materials on the magnetization dynamics of adjacent HM/amorphous FM bilayer is yet to be explored. The magnetization dynamics represented by phenomenological Gilbert damping parameter $\alpha_{eff}$ is mainly contributed by intrinsic SOC of FM and extrinsic effects arising due to two magnon-scattering (TMS), spin pumping, and interfacial spin memory loss (SML). The interfacial SML can arise due to the interfacial SOC (ISOC) that can act as an additional spin sink and enhance the magnetic damping \cite{zhu2019effective,zhu2020origin}. At the same time, there can be various origins of TMS in HM/FM/CL systems. Especially, the magnetic defects or roughness at the interface can scatter the uniform magnon mode of a precessing macrospin due to the presence of non-uniform short wavelength magnons, leading to the TMS \cite{zhu2019effective,zhu2020origin}. These additional extrinsic effects contribute to the magnetization relaxation rate and enhance the effective damping. The reduction of these effects is of primary importance when it comes to energy efficient current induced magnetization switching, magnetization oscillation, magnon propagation, current-induced skyrmion motion etc. Further, the capping layer (CL) employed to protect the FM from oxidation in HM/FM bilayers can also significantly affect the magnetic damping due to additional FM/Capping layer interface contribution to Rashba states, TMS and SML \cite{behera2017capping}. Especially, the capping layer usually gets naturally oxidized in the ambient environment and can act like a sink for spin and/or orbital angular momentum \cite{santos2023inverse, zheng2024enhanced}. 
\par
In this regard, we report the significant modifications of magnetization dynamics and spin-orbital to charge conversion when the $\beta$-W/Co$_{20}$Fe$_{60}$B$_{20}$ (CFB) bilayers are capped by the inorganic (CuO$_x$) and organic (C$_{60}$) capping layers. The C$_{60}$ over layer induces a faster magnetization relaxation, however contrastingly, the spin-orbital pumping induced charge current is found to be enhanced for CuO$_x$ capping. The experimental results are explained by the anti-damping and ORE phenomena associated with CuO$_x$ and the possible TMS, SML induced by the CFB/C$_{60}$ interface.
\section{Experimental Methods}
Two different series of heterostructures with $\beta$-W (10 nm)/CFB (4, 6, 8,10 nm)/CuO$_x$ (3 nm) and $\beta$-W (10 nm)/CFB (4, 6, 8,10 nm)/C$_{60}$ (25 nm) stackings have been fabricated on Si/SiO$_2$ (300 nm) substrates (Fig. \ref{fig:schematics} (a-b)). Along with that the CFB/CuO$_x$ and CFB/C$_{60}$ bilayers were also prepared for the investigation of magnetization dynamics and spin/orbital pumping phenomena. The heterostructures can be read as follows:
\begin{table}
\caption{\label{tab:table1}
Stacking of different heterostructures with their corresponding nomenclatures}
\begin{ruledtabular}
\begin{tabular}{lr}
\multicolumn{1}{c}{\textrm{Stacking}}&
\textrm{Nomenclature}\\
\colrule
Si/SiO$_2$ (300 nm)/$\beta$-W (10 nm)/CFB (4 nm)/CuO$_x$ (3 nm)	& WCF1 \\
Si/SiO$_2$ (300 nm)/$\beta$-W (10 nm)/CFB (6 nm)/ CuO$_x$ (3 nm) &	WCF2 \\
Si/SiO$_2$ (300 nm)/$\beta$-W (10 nm)/CFB (8 nm)/ CuO$_x$ (3 nm) &	WCF3 \\
Si/SiO$_2$ (300 nm)/$\beta$-W (10 nm)/CFB (10 nm)/ CuO$_x$ (3 nm) &	WCF4 \\
Si/SiO$_2$ (300 nm)/$\beta$-W (10 nm)/CFB (4 nm)/C$_{60}$ (25 nm)	& WCFO1 \\
Si/SiO$_2$ (300 nm)/$\beta$-W (10 nm)/CFB (6 nm)/C$_{60}$ (25 nm)	& WCFO2 \\
Si/SiO$_2$ (300 nm)/$\beta$-W (10 nm)/CFB (8 nm)/C$_{60}$ (25 nm)	& WCFO3 \\
Si/SiO$_2$ (300 nm)/$\beta$-W (10 nm)/CFB (10 nm)/C$_{60}$ (25 nm) &	WCFO4 \\
\end{tabular}
\end{ruledtabular}
\end{table}
\newpage
The CFB, $\beta$-W, and Cu layers were grown by DC magnetron sputtering and C$_{60}$ layer was deposited in-situ via effusion cells (Manufactured by EXCEL Instruments, India) at room temperature. The top thin Cu layer oxidizes naturally to form the CuO$_x$ capping for the heterostructures in the WCF series. We have also fabricated 10 and 20 nm Cu thin films on Si/SiO$_2$ to investigate the formation of the natural oxidation of Cu. Before the fabrication of heterostructures, individual thin films of CFB, W, C$_{60}$ and Cu were prepared for thickness calibration and study of magnetic and electrical properties. The base pressure of the sputtering chamber was maintained at $\sim$ 4 $\times$ 10$^{-8}$ mbar prior to the deposition. The structural characterizations of individual thin films and heterostructures were performed by x-ray diffraction (XRD), x-ray reflectivity (XRR) techniques, and Raman spectrscopy. The surface topography was imaged by atomic force microscope (AFM). The superconducting quantum interference device based vibrating sample magnetometer (SQUID-VSM) and magneto-optic Kerr effect (MOKE) based microscope were employed for the static magnetization characterization. The magnetization dynamics was investigated by a lock-in based ferromagnetic resonance (FMR) spectrometer manufactured by NanOsc. The heterostructures were kept in a flip-chip manner on the co-planner waveguide (CPW) and the FMR spectra in 4-17 GHz range were recorded for all the samples. The FMR spectrometer set-up is also equipped with an additional nano voltmeter using which spin-orbital to charge conversion phenomena of all the devices were measured via inverse spin Hall effect (ISHE) with 15 dBm RF power. The contacts were given at the two opposite ends of 3 mm $\times$ 2 mm devices using silver paste to measure the ISHE induced voltage drop across the samples.
\section{Results and Discussion}
\begin{figure*}
\includegraphics[scale=0.6]{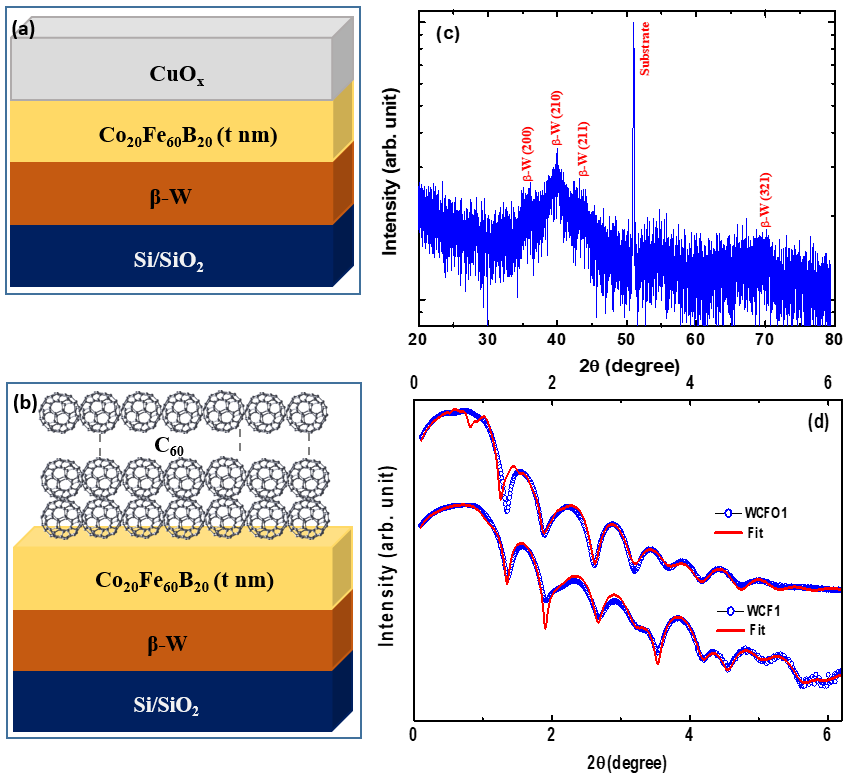}
\caption{\label{fig:schematics}
Schematics of sample stacking for (a) WCF and (b) WCFO series of heterostructures. (c) GIXRD pattern of $\beta$-W (10 nm)/CFB (4 nm)/CuO$_x$ (3 nm) [WCF1] heterostructure and (d) XRR patterns and the corresponding fits with GenX software for $\beta$-W (10 nm)/CFB (4 nm)/CuO$_x$ (3 nm) [WCF1] and $\beta$-W (10 nm)/CFB (4 nm)/C$_{60}$ (25 nm) [WCFO1] heterostructures.}
\end{figure*}

The grazing incidence x-ray diffraction (GIXRD) measurements were performed for all the heterostructures. The GIXRD pattern of WCF1 heterostructure is shown in Fig. \ref{fig:schematics} (c). The presence of (200), (210), (211), and (321) Bragg peaks confirm the stabilization of $\beta$-phase of W. A similar type of GIXRD pattern is also observed for other heterostructures. The x-ray reflectivity (XRR) patterns of WCF1 and WCFO1 heterostructures are shown in Fig. \ref{fig:schematics} (d). 
The XRR data of all the heterostructures are fitted with GenX software and the desired thickness of individual layers are confirmed from the XRR fittings. Further, the details of growth and structural characterizations of $\beta$-W, C$_{60}$, and naturally oxidized Cu (CuO$_x$) are mentioned elsewhere \cite{sahoo2024molecular,sahoo2024efficient}. The nonlinear I-V characteristics along with the presence of Bragg peaks for CuO$_x$ in the GIXRD patterns of 10 and 20 nm Cu/CuO$_x$ films grown on Si/SiO$_2$ (300 nm) confirm the presence of CuO$_x$ \cite{sahoo2024efficient}.   
\begin{figure*}
\includegraphics[scale=0.5]{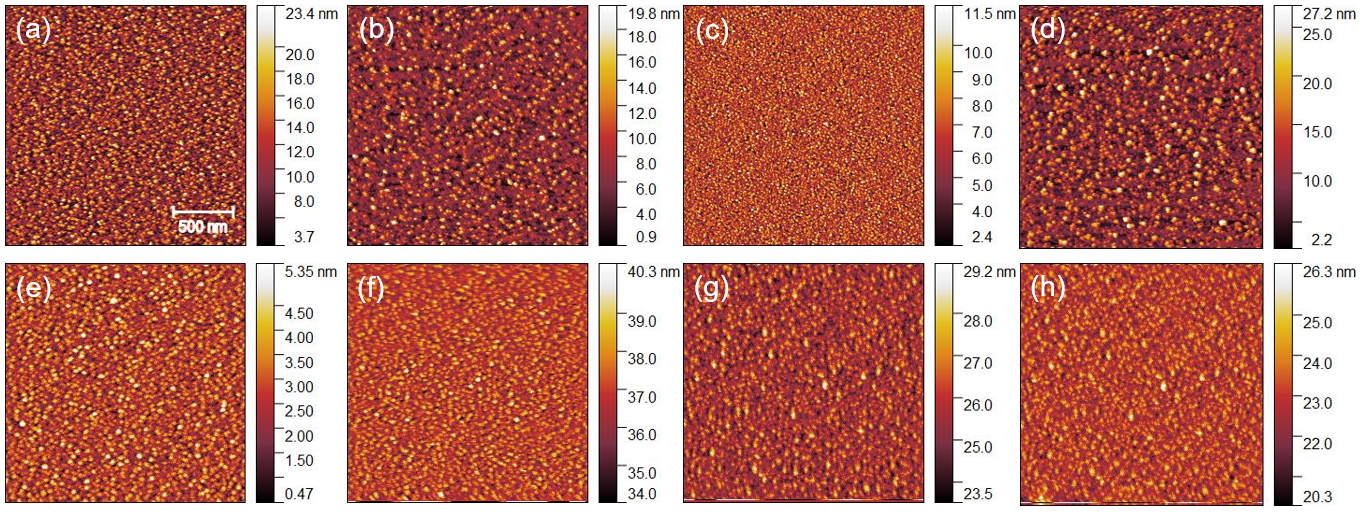}
\caption{\label{fig:AFM} AFM topography of (a) WCF1, (b) WCF2, (c) WCF3, (d) WCF4, (e) WCFO1, (f) WCFO2, (g) WCFO3, and (h) WCFO4 heterostructures. Scale bars correlating the colour to the height are shown to the right of each image.}
\end{figure*}
\par
The surface morphology of all the heterostructures imaged by AFM are presented in Fig. \ref{fig:AFM}. The topography changes for different heterostructures in the WCF series. The RMS roughness for WCF1, WCF2, WCF3, and WCF4 heterostructures are found to be $\sim$ 3.68 nm,$\sim$ 3.02 nm, $\sim$ 1.5 nm, and $\sim$ 4.2 nm, respectively. Although the thickness of CuO$_x$ capping layer is same for all the heterostructures in the WCF series, the roughness is found to be random. This might be due to the different nature of natural oxidation of Cu as it is occurring in the ambient environment in an uncontrolled manner. Such type of oxidation can play an important role in governing the magnetization dynamics properties of these heterostructures. We have also performed the AFM imaging of 6 and 10 nm thick CFB films (Fig. S1 of Supplemental Material \cite{Supplementary}). The RMS roughness is found to be similar ($\sim$ 0.6 nm) for both the films, further inferring the change in roughness in WCF series is due to the top CuO$_x$ layer. Whereas, the topography of the heterostructures in the WCFO series when capped by the organic C$_{60}$ does not change much. The RMS roughness of all the heterostructures in WCFO series are found to similar ($\sim$ 0.8 to $\sim$ 1 nm), in contrast to the heterostructures in WCF series.
\par
\begin{figure*}
\includegraphics[scale=0.6]{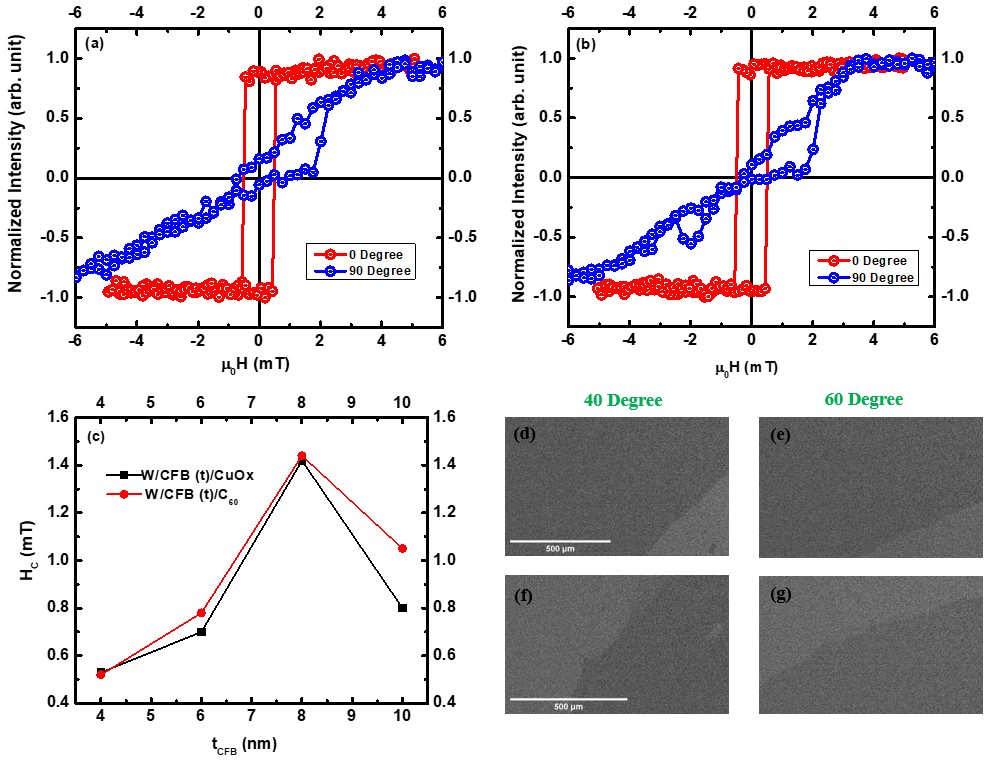}
\caption{\label{fig:AMOKE} Hysteresis loops of (a) $\beta$-W (10 nm)/CFB (4 nm)/CuO$_x$ (3 nm) [WCF1] and (b) $\beta$-W (10 nm)/CFB (4 nm)/C$_{60}$ (25 nm) [WCFO1] heterostructures measured along easy and hard axes, (c) CFB thickness dependent $H_c$ of various heterostructures, and magnetic domain images of (d-e) WCF1 and (f-g) WCFO1 heterostructures.}
\end{figure*}

The angle-dependent in-plane hysteresis loops were traced for all the heterostructures via MOKE microscopy. The magnetization reversals for WCF1 and WCFO1 heterostructures along the easy and hard axes are presented in Fig. \ref{fig:AMOKE} (a-b). Both the heterostructures exhibit the presence of uniaxial magnetic anisotropy, which can be attributed to the oblique incident sputtered growth of the FM layer. A similar type of magnetization reversal has also been observed for other heterostructures capped with CuO$_x$ and C$_{60}$. 
A significant change in the coercive field ($H_c$) is not evident when the capping layer is changed from CuO$_x$ to C$_{60}$. The capping of organic C$_{60}$ on 3d transition ferromagnetic metals usually enhances the magneto-crystalline anisotropy (MCA) energy and consequently, the coercive field owing to interfacial hybridization \cite{mallik2019enhanced,pandey2023perspective}. However, the modification of magnetic anisotropy energy is highly dependent on the crystallinity, orbital orientation, exchange correlation length, etc \cite{pandey2023perspective}. The as-deposited CFB is amorphous in nature and usually composed of nano-crystalline grains in 1-10 nm range \cite{chen2010effect}. The presence of relatively smaller in size and large number of grains can statistically average out the magneto-crystalline anisotropy and reduce the effective anisotropy energy (K$_{eff}$) in CFB \cite{chen2010effect,gupta2022role}. The reduction in K$_{eff}$ can enhance the ferromagnetic exchange correlation length as per the following equation \cite{chen2010effect}.
\begin{equation}
    L_{ex} = \pi (\frac{A_{eff}}{K_{eff}})^{\frac{1}{2}}
    \label{eq:ExcLen}
\end{equation}
Here, $A_{eff}$ represents the effective exchange stiffness and $L_{ex}$ represents the range of effective exchange interaction. This is also evident as the hysteresis loops with high squareness along the easy axes for both the heterostructures (Fig \ref{fig:AMOKE} (a-b)). Further, a relatively larger magnetic domain is observed for CFB (Fig. \ref{fig:AMOKE} (d-g)) in all the heterostructures compared to the previous report \cite{sharangi2021spinterface}. This also signifies the presence of larger exchange correlation length. As the organic over layer primarily modifies the MCA, the amorphous nature of CFB can explain the similar $H_c$ for both inorganic and organic capping layers as well as the soft magnetic nature of WCF1 and WCFO1. The ferromagnets with low $H_c$ are quite important from the spintronics application point-of-view. Interestingly, the $H_c$ of the hysteresis loops measured along the easy axes for different heterostructures increases gradually with the increase in CFB thickness (Fig. \ref{fig:AMOKE} (c)). The gradual increase of $H_c$ with thickness is observed for both inorganic and organic capping layers. This reflects the systematic growth and nucleation of different layers in both the series of heterostructures considered for the present study. Previously, a similar type of thickness-dependent $H_c$ has been observed for as-deposited amorphous CFB with higher thicknesses \cite{chen2010effect,chen2012magnetic}. This has been attributed to the increase in grain size of the nano-crystallites of CFB. The larger grains with a reduction in number of grain might not average out the local MCA. Hence, a larger $K_{eff}$ and consequently, a larger $H_c$ can be expected with the increase in CFB thickness. The decrease in $H_c$ for heterostructures with thicker 10 nm CFB could be due to the decrease in pinning of the domain wall motion during the magnetization reversal as the nucleation of grains are expected to be saturated for thicker CFB. 
\par
\begin{figure*}
\includegraphics[scale=0.6]{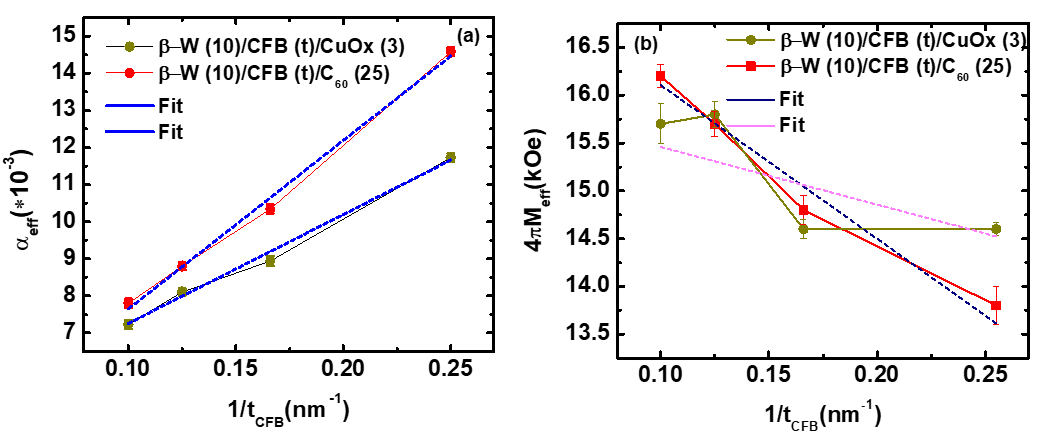}
\caption{\label{fig:FMR2} 1/t$_{CFB}$ dependent (a) $\alpha_{eff}$ and (b) 4$\pi M_{eff}$ for various heterostructures and the corresponding linear fits to equation \ref{eq:alpha} and \ref{eq:Kittel}.}
\end{figure*}

The magnetization dynamics of all the heterostructures in both the WCF and WCFO series were investigated by lock-in based FMR technique. The heterostructures were placed in a flip-chip manner on top of the co-planner waveguide (CPW) and the FMR spectra were recorded in the 4-17 GHz range. The field swept FMR spectra at different resonance frequencies for WCF1 and WCFO1 heterostructures are shown in Fig. S2 (Supplemental Material \cite{Supplementary}). A similar type of FMR spectra were recorded for all the heterostructures. Each spectrum was fitted by the derivative of symmetric and anti-symmetric components of the Lorentzian function\cite{sahoo2024molecular}:
\begin{equation}
    FMR  Signal = K_1 \frac{4(\Delta H)(H-H_{res})}{[(\Delta H)^2 + 4 (H - H_{res})^2]^2} - K_2 \frac{(\Delta H)^2 - 4(H-H_{res})^2}{[(\Delta H)^2 + 4 (H - H_{res})^2]^2} + Offset,
    \label{eq:FMR}
\end{equation}
where $K_1$ and $K_2$ are the antisymmetric and symmetric absorption co-efficients, respectively.
\begin{figure*}
\includegraphics[scale=0.6]{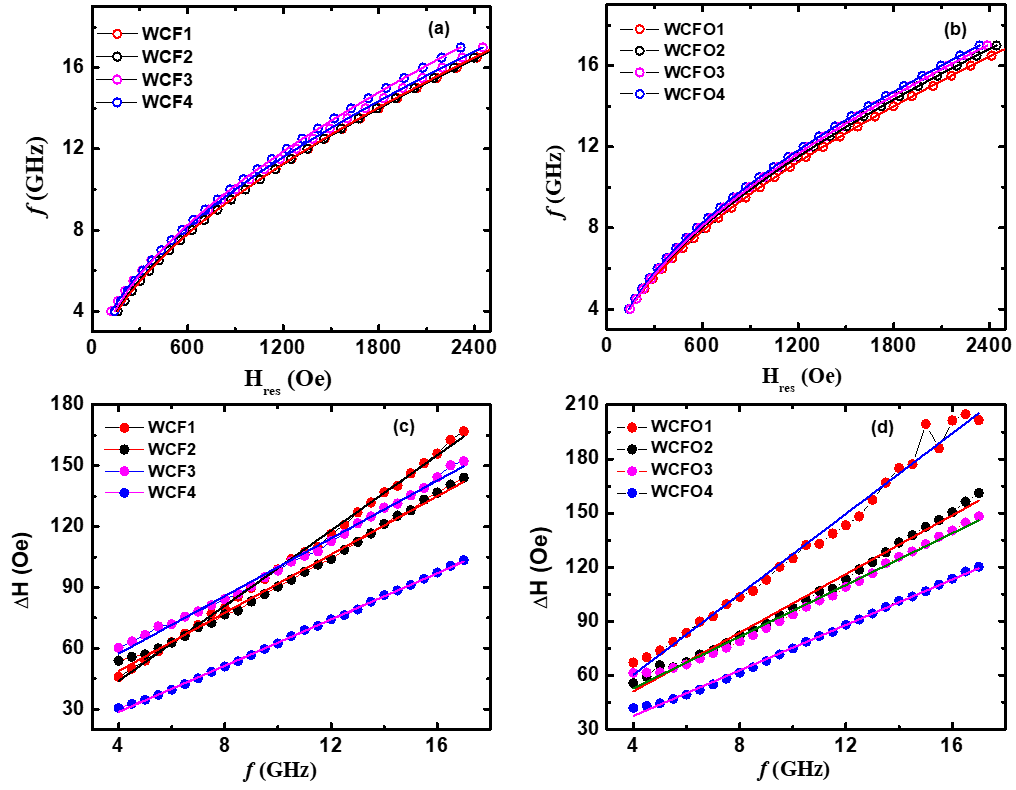}
\caption{\label{fig:FMR} (a-b) Frequency ($f$) vs resonance field ($H_{res}$) and (c-d) linewidth ($\Delta H$) vs frequency ($f$) behaviour for various heterostructures. The solid lines are the best fits to equations \ref{eq:Kittel} and \ref{eq:damping}.}
\end{figure*}
The resonance field ($H_{res}$) and linewidth ($\Delta H$) extracted for various resonance frequencies from the Lorentzian fit of the field dependent FMR absorption are shown in Fig. \ref{fig:FMR}. The $H_{res}$ dependent $f$ of different heterostructures are plotted in Fig. \ref{fig:FMR} (a-b). The $f$ vs $H_{res}$ plots are fitted by using the Kittel’s equation \cite{sahoo2024molecular}:
\begin{equation}
    f = \frac{\gamma}{2 \pi} \sqrt{(H_K + H_{res})(H_K +H_{res}+4\pi M_{eff})},
    \label{eq:Kittel}
\end{equation}
where$$4\pi M_{eff} = 4\pi M_s + \frac{2K_s}{M_s t_{FM}}$$
and $H_K$, $K_s$, and $t_{FM}$ are the anisotropy field, interface magnetic anisotropy energy density, and the thickness of FM, respectively. Here, $\gamma$ is the gyromagnetic ratio and 4$\pi M_{eff}$ represents the effective magnetization. The 4$\pi M_{eff}$ extracted from the fitting gives similar values as compared with the saturation magnetization value (4$\pi M_{s}$) calculated from the SQUID-VSM (See the supplemental Material \cite{Supplementary}). The magnetic Gilbert damping which encompasses pivotal information regarding magnetization relaxation, spin wave propagation, spin-pumping into the adjacent non-magnetic layers is investigated from the resonance frequency dependent FMR linewidth behavior (Fig. \ref{fig:FMR} (c-d)). A linear dependency of $\Delta H$ on resonance frequency is evident for all the heterostructures in both WCF and WCFO series. The $\Delta H$ vs $f$  plots are fitted by the following equation to separate the intrinsic and extrinsic contribution to the precessional damping \cite{sahoo2024molecular}:
\begin{equation}
    \Delta H = \Delta H_0 + \frac{4 \pi \alpha_{eff}}{\gamma}f
    \label{eq:damping}
\end{equation}
Here, $\Delta H_0$ is known as the linewidth broadening caused by the sample imperfections representing the extrinsic contribution. The $\alpha_{eff}$ represents the intrinsic contribution to the damping and also known as effective Gilbert damping. The $\alpha_{eff}$ of different heterostructures were evaluated from the linear fits of $\Delta H$ vs $f$ plots using equation \ref{eq:damping}. The 1/t$_{CFB}$ dependent $\alpha_{eff}$ of different heterostructures from both the series are shown in Fig. \ref{fig:FMR2} (a). Interestingly, the $\alpha_{eff}$  for WCFO series are found to be larger compared to that with the heterostructures in the WCF series. Especially, the enhancement is more prominent for the heterostructures with thinner CFB layers. This infers the significant modification in magnetization dynamics when the $\beta$-W/CFB bilayers are capped by inorganic and organic layers. The enhancement of Gilbert damping when the HM/FM bilayers are capped by C$_{60}$ can have different origins. The additional spin-pumping into organic layer, two magnon scattering caused by the interfacial SOC and magnetic roughness at the CFB/C$_{60}$ interface or interfacial spin memory loss could be the reason for this significant enhancement. Although the static magnetization properties of the heterostructures capped by C$_{60}$ and CuO$_{x}$ remains similar, the magnetization relaxation phenomenon under microwave excitation presents a clear difference. The $\alpha_{eff}$ for both the CuO$_{x}$ and C$_{60}$ capping increases linearly with 1/t$_{CFB}$ and can be well fitted with the following equation \cite{behera2017capping}:
\begin{equation}
    \alpha_{eff} = \alpha_{CFB} + g_{eff}^{\uparrow\downarrow} \frac{g \mu_B}{4 \pi M_s t_{CFB}}
    \label{eq:alpha}
\end{equation}
 where $\alpha_{eff}$ is the intrinsic damping of CFB layer. g, $\mu_B$ and t$_{CFB}$ are the Landé g factor (2.1), Bohr’s Magnetron and thickness of the CFB layer, respectively.
 The slopes of the linear fit for WCF and WCFO series are found to be significantly different and can influence the spin-orbital to charge interconversion phenomenon in $\beta$-W/CFB/CL heterostructures. The g$_{eff}^{\uparrow\downarrow}$, which is the real part of spin mixing conductance, were calculated for WCF and WCFO series by considering the $M_s$ value $\sim$ 1200 emu/cc, measured by SQUID VSM (See Supplemental Material (Fig. S3) \cite{Supplementary}).  The g$_{eff}^{\uparrow\downarrow}$ values for WCF and WCFO series are found to be 2.2$\times$10$^{19}$ m$^{-2}$ and 3.5$\times$10$^{19}$ m$^{-2}$, respectively. Here, it is important note that the heterostructures investigated in this work comprise CFB thickness in 4-10 nm range. The $\alpha_{eff}$ vs 1/t$_{CFB}$ behavior may deviate from the linear dependency for thinner CFB layers due to the contributions from TMS and SML \cite{yoshii2022significant}. However, the TMS and SML are not expected to play a significant role for thicker CFB as in our case. But the organic CL could induce local orbital hybridization and hence, the modification of interface electronic structure and local magnetic anisotropic energy. This could enhance the TMS as it is usually proportional to square of $\frac{2K_s}{M_s}$ \cite{yoshii2022significant}. The 4$\pi M_{eff}$ vs 1/t$_{CFB}$ behavior for all the heterostructures are shown in Fig. \ref{fig:FMR2} (b). The 4$\pi M_{eff}$ varies linearly with 1/t$_{CFB}$ for WCFO series, whereas the linear behavior is absent for WCF series. This indicates the $K_s$, which represents the interfacial magnetic anisotropy energy density for both types of interfaces on either side of FM layer, may not be same for all the heterostructures in WCF series. The Cu capping in WCF series gets naturally oxidized to form CuO$_x$ and the oxidation level could be different in different heterostructures as it is not controlled experimentally. This behavior is also consistent with randomness in the surface topographic images observed for different heterostructures in WCF series. Hence, the interfacial anisotropy in CFB/CuO$_x$ could be modulated for different heterostructures in WCF series as the 3 nm Cu is expected to be completely oxidized \cite{sahoo2024efficient, santos2024exploring, ding2024mitigation}. Whereas, the thicker 25 nm C$_{60}$ capping presents a similar FM/C$_{60}$ interface for all the heterostructures in WCFO series and hence, the similar $K_s$  for all the heterostructures with organic capping. Nevertheless, a rough estimation of slope from the linear fit of 4$\pi M_{eff}$ vs 1/t$_{CFB}$ behavior in Fig. \ref{fig:FMR2} (b) can shed light on possible origin of enhanced Gilbert damping for WCFO series. As it can be seen in Fig. \ref{fig:FMR2} (b), the slope (and hence the $\frac{2K_s}{M_s}$   (from equation \ref{eq:Kittel})) value is larger for WCFO series compared to the WCF series. This infers a possible local short-range interfacial hybridization upon C$_{60}$ capping, which could possibly modify the $K_s$ and induce relatively faster magnetization precession via magnon-magnon scattering\cite{behera2017two}. Further, the interfacial spin memory loss due to ISOC at the CFB/C$_{60}$ interface could also enhance the magnetic damping as the C$_{60}$ is predicted to possess curvature induced SOC \cite{sun2016inverse}. 
\begin{figure*}
\includegraphics[scale=0.45]{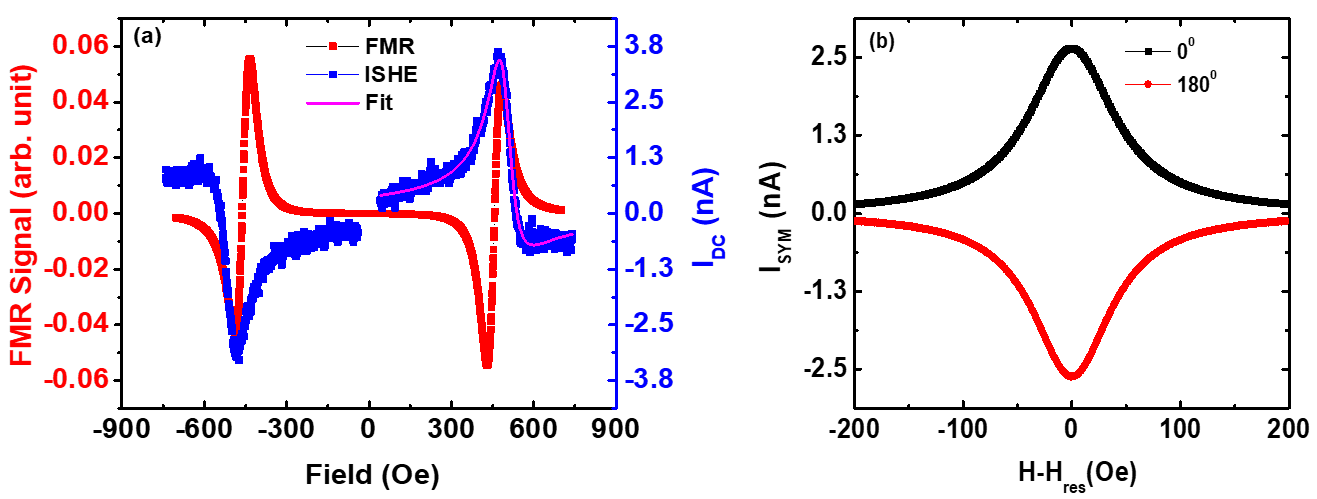}
\caption{\label{fig:ISHE1} (a) Magnetic field dependent FMR spectrum and measured DC current ($I_{DC}$) and (b) symmetric components of DC current ($I_{SYM}$) for $\beta$-W (10 nm)/CFB (4 nm)/CuO$_x$ (3 nm) [WCF1] heterostructure measured at 7 GHz frequency.}
\end{figure*}
\begin{figure*}
\includegraphics[scale=0.6]{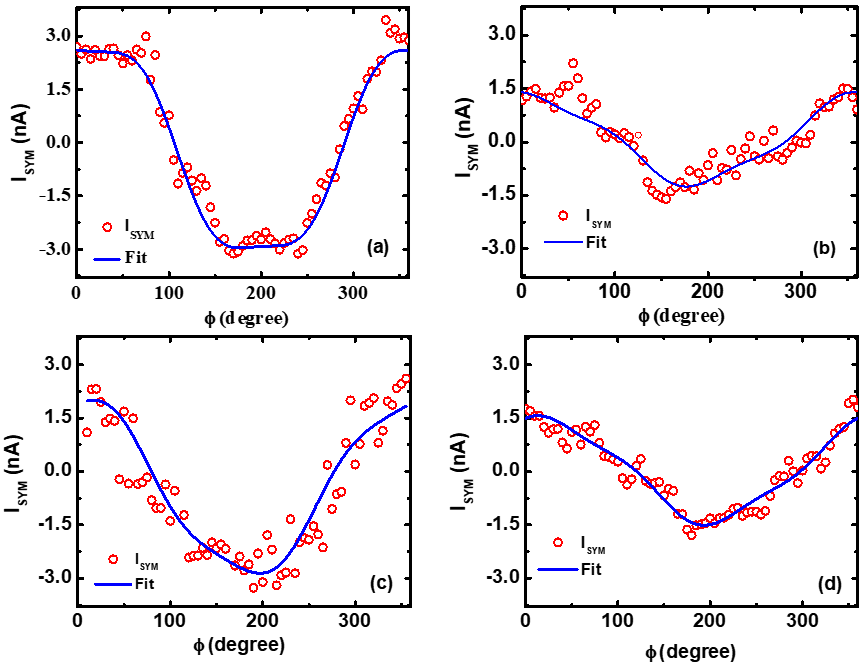}
\caption{\label{fig:ISHE2} Angle dependent symmetric components of DC current ($I_{SYM}$) for (a) WCF1, (b) WCFO1, (c) WCF2, and (d) WCFO2 heterostructures measured at 7 GHz frequency.}
\end{figure*}
 \par
 In order to further understand the magnetization dynamics of the heterostructures, the spin/orbital to charge conversion phenomenon of all the samples in WCF and WCFO series were investigated. The measurements were performed from $\phi \sim$ 0$^{\circ}$ to $\phi \sim$ 360$^{\circ}$, where $\phi$ represents the angle between the measured voltage direction and perpendicular direction to applied magnetic field during ferromagnetic resonance. The field swept FMR and corresponding measured DC current across the WCF1 sample is shown in Fig. \ref{fig:ISHE1} (a). The sign of the measured DC current ($I_{DC}$ $\sim$ $V_{MEAS}$/$R$: $V_{MEAS}$ is the measured DC voltage and $R$ is the device resistance) gets reversed for opposite external field direction, inferring the spin-pumping mechanism. A similar type of $I_{DC}$ vs $H$ pattern is observed for all the heterostructures in WCF and WCFO series. The $I_{DC}$ vs $H$ plots are fitted by the following Lorentzian function to separate the symmetric ($I_{SYM}$) and asymmetric ($I_{ASYM}$) components of the measured DC currents \cite{sahoo2024molecular}:
\begin{equation}
    I_{DC} = I_{SYM} \frac{(\Delta H)^2}{(\Delta H)^2 + (H - H_{res})^2} + I_{ASYM} \frac{(\Delta H) (H-H_{res})}{(\Delta H)^2 + (H - H_{res})^2} 
    \label{eq:ISHE}
\end{equation} 
The $I_{SYM}$ around the resonance frequency extracted for $\phi \sim$ 0$^{\circ}$ also reverses the sign for $\phi \sim$ 180$^{\circ}$ (Fig. \ref{fig:ISHE1} (b)), as expected for typical ISHE measurements. The symmetric voltage component ($V_{SYM}$) comprises the spin-pumping induced DC voltage $V_{SP}$ along with spin rectification effects arising due to anisotropic magnetoresistance (AMR) and anomalous Hall effect (AHE). Further, the $\beta$-W has a negative spin Hall angle, whereas the spin/orbital Hall angle of C$_{60}$ and CuO$_x$ are found to be positive \cite{sahoo2024efficient, santos2024exploring, sun2016inverse, santos2023inverse}. Hence, the $I_{DC}$ at the $\beta$-W/CFB interface and CFB/CL interface are expected to be added up according to the symmetry. The angle dependent $I_{SYM}$ plots for WCF1, WCFO1, WCF2, and WCFO2 heterostructures are shown in Fig. \ref{fig:ISHE2} (a-d).  The data are fitted with the following equation to exclude the spin rectification effects and evaluate the $I_{SP}$ \cite{sahoo2024molecular}:
\begin{equation}
\begin{split}
    I_{SYM} = I_{SP} Cos^3 (\phi) +  I_{AHE} Cos(\phi) Cos(\theta) \\ + I_{SYM}^{AMR\perp} Cos (2\phi) Cos(\phi) + I_{SYM}^{AMR\parallel} Sin (2\phi) Cos(\phi)
\end{split}
\end{equation}
Here, $\theta$ is the phase between RF electric field and magnetic field in the medium. $I_{AHE}$, $I_{SYM}^{AMR\perp}$, $I_{SYM}^{AMR\parallel}$ are the charge current arising due to AHE, perpendicular component of current arising due to AMR and parallel component of current arising due to AMR, respectively. A similar type of fitting of $I_{SYM}$ vs $\phi$ plots have also been performed for other heterostructures to evaluate the $I_{SP}$ ($I_{SP}$ $\sim$ $\frac{V_{SP}}{R}$, where $R$ is the device resistance).
\begin{figure*}
\includegraphics[scale=0.55]{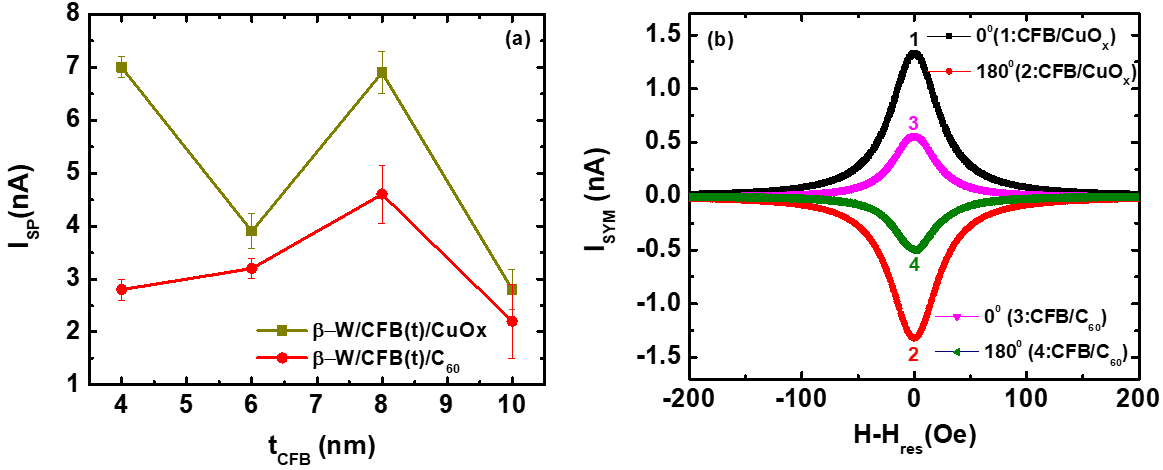}
\caption{\label{fig:ISHE3} (a) CFB thickness dependent spin pumping current ($I_{SP}$) and (b) symmetric components of DC current ($I_{SYM}$) for CFB (7nm)/CuO$_x$ (3 nm) and CFB (7nm)/C$_{60}$ (25 nm) bilayers.}
\end{figure*}
\par
The $I_{SP}$ for all the heterostructures in WCF and WCFO series are plotted in Fig. \ref{fig:ISHE3} (a). Interestingly, the $I_{SP}$ values for the heterostructures with CuO$_x$ capping is found to be larger compared to that with organic C$_{60}$ capping. This trend is of opposite nature to that of $\alpha_{eff}$, where the magnetization relaxation is found to be faster for the organic C$_{60}$ capping. To further understand this anti-correlation effect, we measured the spin/orbital to charge conversion of CFB (7nm)/CuO$_x$ (3 nm) and CFB (7nm)/C$_{60}$ (25 nm) heterostructures. The $I_{SYM}$ for both the heterostructures reverse sign when the $\phi$ was changed from 0$^\circ$ to 180$^\circ$ (Fig. \ref{fig:ISHE3} (b)), confirming the spin/orbital pumping in both the bilayers with non-magnetic low SOC over layers. The evolution of spin-orbital to charge current interconversion in FM/CuO$_x$ bilayer has been attributed previously to orbital Rashba effect \cite{ding2022observation, kim2023oxide}. However, in those cases different FMs, like NiFe, Co were employed. Our experiments show that the amorphous CFB layer can also act as an orbital angular momentum source for orbital pumping. At the same time, the realization of DC current in CFB (7nm)/C$_{60}$ (25 nm) bilayer under FMR conditions is quite interesting. Previously, it has been detected by the pulsed-ISHE technique with NiFe/C$_{60}$ bilayer and is attributed to the curvature induced SOC in C$_{60}$ \cite{sun2016inverse}. However, the detailed theoretical investigation is required to unravel the exact origin of spin/orbital pumping induced charge current in these low Z materials. Further, the measured charged current for CFB/CuO$_x$ bilayer is found to be quite larger compared to that with CFB/C$_{60}$ bilayer. This can explain the larger $I_{SP}$ observed for heterostructures capped with CuOx in WCF series compared to the heterostructures in WCFO series. The $I_{SP}$ exhibits an anomalous behavior with increase in CFB thickness for WCF series which can be correlated with anomalous change of $K_s$ with CFB thickness as observed in Fig. \ref{fig:FMR2} (b). As discussed earlier, the natural oxidation of Cu could affect the interface magnetic anisotropy at CFB/CuO$_x$ interface in an irregular manner and hence, the $K_s$ can change accordingly. This could modify the CFB/CuO$_x$ interface transparency for injection of orbital current and consequently, can tune the $I_{SP}$. The spin-orbital to charge conversion results also unravel the fact that the organic C$_{60}$ capping only enhances the magnetic damping without much increasing the DC current in the stack compared to inorganic CuO$_x$ capping. This infers the relatively faster magnon relaxation in WCFO series could be facilitated by the magnon-magnon scattering rather than the spin/orbital pumping. In addition, the SML at the CFB/C$_{60}$ interface can also play a role in enhancing the damping value as the C$_{60}$ layer offers the curvature induced SOC leading the metal/organic interface to act as an additional spin sink \cite{sun2016inverse}. Here, it is also important to note that the Rashba like states at CFB/CuO$_x$ interface could lead to the non-equilibrium orbital angular momentum (OAM) accumulation near the interface in WCF series. The accumulated OAM could be converted to charge current via IOREE and could also induce anti-damping like torque on magnetization resulting a reduction in effective damping value.  Simillar type of anti-damping like torque and hence, the reduced damping value has been previously reported in NiFe/TaO$_x$ bilayer \cite{behera2016anomalous}. Thus, the anti-damping effect at CFB/CuO$_x$ interface cannot be completely ignored while comparing the effective Gilbert damping of heterostructures in WCF and WCFO series. Our detailed experiments show that the CL can act as an additional source for effective damping control as well as generation of spin/orbital pumping induced charge current. It also sheds light on the important contribution of Rashba interface, TMS, and SML in evaluation of technologically important Gilbert damping parameter and necessity of spin-orbital to charge interconversion study to interpret magnetization relaxation results even for thicker FM films rather than merely investigating the magnetic damping from FMR. Especially, the CuO$_x$ over layer can serve the purpose for achieving efficient spin-orbital to charge interconversion without much of increasing the magnetic Gilbert damping and hence, provides a suitable alternative for the development of power efficient spin-orbitronics devices.        
\section{Conclusion}
In conclusion, we have experimentally investigated the effect of organic C$_{60}$ and inorganic CuO$_x$ capping layer on the magnetization dynamics in HM/FM/CL heterostructures. At first, the static magnetic properties measured by the magnetometry techniques reveal the similar behavior for both the capping layers. However, the magnetization relaxation process is found to be faster in the heterostructures with C$_{60}$ capping compared to the CuOx over layer. On contrary, the spin/orbital-to-charge conversion phenomenon gets enhanced with top CuO$_x$ layer in $\beta$-W/CoFeB/CL heterostructures. The results are ascribed to the possible interfacial phenomenon, like orbital Rashba effect, magnon-magnon scattering and spin memory loss at CoFeB/CL interface. Our findings unravel the importance of low SOC materials in controlling the critical magnetization dynamics parameters for efficient spin-orbitronics applications. Moreover, the surface oxidized Cu can facilitate an efficient spin-orbital to charge conversion without much enhancing the magnetic damping, inferring the harnessment of OAM could be path forward for fabrication of low power spin-orbitronics devices. 

\section{Acknowledgements}
We acknowledge the Department of Atomic Energy (DAE), the Department of Science and Technology (DST) of the Government of India, and SERB project CRG/2021/001245. A.S. acknowledges the DST-National Postdoctoral Fellowship in Nano Science and Technology.  We are also thankful to the Center for Interdisciplinary Sciences, NISER for providing Raman spectroscopy measurement facility.
\nocite{*}

\bibliography{apssamp}% Produces the bibliography via BibTeX.

\begin{thebibliography}{30}
\bibitem [{Sup()}]{Supplementary}%
  \BibitemOpen
  \bibfield  {title} {\bibinfo {title} {See \uppercase{S}upplemental \uppercase{M}aterial at [url will be inserted by publisher] for additional structural analysis and magnetic characterization},\ }\href@noop {} {\ }\BibitemShut {NoStop}%
\end{thebibliography}

\end{document}